# Resolution enhancement in in-line holography by numerical compensation of vibrations

Tatiana Latychevskaia* and Hans-Werner Fink
*Physics Institute, University of Zurich, Winterthurerstrasse 190, 8057 Zurich, Switzerland*
*tatiana@physik.uzh.ch*

**Abstract:** Mechanical vibrations of components of the optical system is one of the sources of blurring of interference pattern in coherent imaging systems. The problem is especially important in holography where the resolution of the reconstructed objects depends on the effective size of the hologram, that is on the extent of the interference pattern, and on the contrast of the interference fringes. We discuss the mathematical relation between the vibrations, the hologram contrast and the reconstructed object. We show how vibrations can be post-filtered out from the hologram or from the reconstructed object assuming a Gaussian distribution of the vibrations. We also provide a numerical example of compensation for directional motion blur. We demonstrate our approach for light optical and electron holograms, acquired with both, plane- as well as spherical-waves. As a result of such hologram deblurring, the resolution of the reconstructed objects is enhanced by almost a factor of 2. We believe that our approach opens up a new venue of post-experimental resolution enhancement in in-line holography by adapting the rich database/catalogue of motion deblurring algorithms developed for photography and image restoration applications.

## 1. Introduction

The diffraction-limited resolution of an in-line hologram is defined by the wavelength of the employed radiation $\lambda$ and by the numerical aperture of the setup NA as $R_{NA} = \frac{\lambda}{2NA}$ [1-3], where the numerical aperture is defined by the angular extent of the interference pattern. Many factors can reduce the contrast and the extent of the interference pattern and thus diminish the resolution: noise, partial coherence, limited extent of the reference wave, mechanical vibrations etc. For example, partial coherence leads to decreased contrast of an interference pattern which in turn leads to a blurred reconstruction of the object. Experimental noise in the intensity distribution often buries the fine interference fringes which carry high-resolution information. Vibrations of optical elements also reduce the contrast of the interference pattern. Vibration spectra of some optical elements have previously been measured [4-5]. In this work we address the effect of vibrations and show how this effect can be to some extent compensated by an appropriate post-experimental analysis of holograms.

In in-line holography with plane waves, a shift of the object is linearly transformed into a shift of the entire interference pattern on the detector. In in-line holography with a divergent spherical wave (also called Gabor holography) [6-7], the situation is even more severe, because a shift of either the source or the object both lead to a shift of the interference pattern on the detector. In-line holography with a divergent wave offers a magnified imaging of objects, whereby the magnification is determined by the ratio of source-to-detector to source-to sample distance. The shorter the source-to sample distance the higher is the magnification and hence a higher resolution can be achieved. However as the distance between source and sample gets shorter, the shifts of the hologram due to source or object vibrations are getting larger due to the higher magnification. This in turn blurs out the fine interference fringes and reduces the potentially achievable high resolution.

Experimentally, to compensate for vibrations, a short time acquisition sequence of hologram can be recorded [8] and subsequently the holograms can be aligned by cross-correlation [9] and averaged. Such approach can be successfully applied for a systematic drift of the sample and/or when the acquisition time is much shorter that the period of vibrations. For very fast vibrations, much faster than a single hologram acquisition time, other methods must be designed, which is the subject of this study.

Image deblurring techniques are known in motion-deblurring photo-image analysis. A series of works by Stroke et al. initiated techniques named "holographic deblurring" which employed a recorded hologram of a complex-valued transfer function of the optical system as a deconvolution filter. However, despite the name "holographic deblurring", no holograms are directly involved [10-11]. Deblurring techniques [12-16] have so far never been applied to deblur holograms themselves. One of the reasons might be that deblurring techniques typically require a known motion path or the motion path is recovered during the deblurring procedure. In the latter case, the possible motion path should be limited. In a typical situation of holographic imaging influenced by vibrations, the space of possible motion paths is very large. We show that deblurring can to some extent suppress the effect of vibrations and thus enhance the extent and the contrast of the interference pattern fringes which in turn enhances the resolution of the reconstructed objects.

## 2. Principle

A hologram can be described by $H_0(X,Y)$ where $(X,Y)$ are the coordinates in the detector plane. A lateral shift of the object (or of the source in the case of in-line holography with a spherical wave) by $(\Delta x, \Delta y)$ results in the distribution of the hologram described by $H_0(X+\Delta X, Y+\Delta Y)$ which can mathematically be presented as:

$$H_0(X+\Delta X, Y+\Delta Y) = H_0(X,Y) \otimes \delta(X+\Delta X, Y+\Delta Y), \tag{1}$$

where $\otimes$ denotes the convolution and $\delta(X,Y)$ is the delta-function. Next, we assume that the object (respectively the source) is moving or vibrating laterally (that is in the $(x,y)$-plane) during the acquisition of the hologram. The distribution of the acquired hologram which accumulates all the shifts can be represented as:

$$H(X,Y) = H_0(X,Y) \otimes \sum_i \delta(X+\Delta X_i, Y+\Delta Y_i). \tag{2}$$

The last term in Eq. (2), $\sum_i \left[\delta(X+\Delta X_i, Y+\Delta Y_i)\right]$ is a sum over all possible displacement of the interference pattern, which we name the *vibration function*:

$$V(X,Y) = \sum_i \left[\delta(\Delta X_i, \Delta Y_i)\right]. \tag{3}$$

The vibration function is a sum over all shifts of the interference pattern which is equivalent to a sum over all shifts of the object during the vibrations. This function is called the *blur kernel* or PSF in motion deblurring methods [12-16]. However, we would like to emphasize that using the term PSF would be highly confusing in our case. PSF in our case is a hologram of a point object which is typically a distribution of concentric rings; it is not a position of the object that can be attributed to the motion path. By combining Eqs. (2) and (3) the blurred hologram can be represented as

$$H(X,Y) = H_0(X,Y) \otimes V(X,Y). \tag{4}$$

For simplicity, we further consider in-line holograms acquired with plane waves, whereby $V(x,y) = V(X,Y)$. The obtained results can readily also be applied to in-line holograms acquired with spherical waves, as it has been demonstrated that in the paraxial approximation, the distribution of the holograms acquired with plane and spherical waves are only different by the magnification factor [17]. Provided the vibration function $V(x,y)$ is known, the vibration-free hologram $H_0(X,Y)$ can be obtained by deconvolution:

$$H_0(X,Y) = \text{FT}^{-1}\left\{\frac{\text{FT}[H(X,Y)]}{\text{FT}[V(X,Y)]}\right\} \approx \text{FT}^{-1}\left\{\frac{\text{FT}[H(X,Y)](\text{FT}[V(X,Y)])^*}{|\text{FT}[V(X,Y)]|^2 + \beta}\right\} \quad (5)$$

whereby $\beta$ is a small positive constant added in the denominator to avoid division by small values, FT and FT$^{-1}$ denote the direct and the inverse Fourier transforms, correspondingly. In some cases $\text{FT}[V(X,Y)]$ can be calculated analytically. For example, in the absence of a systematic drift in a specific direction, the vibration function $V(x,y)$ can be assumed to have a Gaussian distribution. Consequently, its Fourier transform, $\text{FT}[V(X,Y)]$, is also a Gaussian distribution which can be readily calculated analytically. We introduce the deconvolution function as:

$$D(\mu,\nu) = \frac{1}{\text{FT}[V(X,Y)]} \quad (6)$$

whereby $(\mu,\nu)$ are the coordinates in the Fourier domain, and the Fourier transform is defined as

$$u(\mu,\nu) = \iint U(X,Y)\exp[-i(X\mu + Y\nu)]\,dXdY. \quad (7)$$

The deconvolution function $D(\mu,\nu)$ reaches its minimal value in the center of the Fourier domain (at the zero frequencies, $D(0,0)$). $D(\mu,\nu)$ should be normalized by division with its minimal value of the amplitude, so that $D(0,0) = 1$. This ensures that the intensity of the constant background (or the constant reference wave) is not altered by the deblurring procedure.

This deblurring method has the following analogy to a coherent diffractive imaging (CDI) experiment [18]. In CDI, like in a crystallographic experiment, the recorded diffraction pattern is insensitive to lateral shifts of the object. This is because a lateral shift of the object results in additional linear phase distribution in the far-field (detector plane) which is lost during the acquisition of just the intensity of the wave. Thus, a lateral movement of the object does not change the diffraction pattern. This is why CDI allows acquiring diffraction patterns at the highest possible resolution. The Fourier transform of a shifted hologram is given by a product of the Fourier transform of original hologram and the linear phase distribution caused by the shift. When several such shifted holograms are added, the Fourier transform of their sum is a product of the Fourier transform of the original hologram and the sum of the phase terms caused by the shifts. The last term thus modulates the Fourier transform of the original hologram. When this modulation term can be removed, the Fourier transform of the original hologram will emerge.

## 3. Simulated results

Simulated results are shown in Fig. 1. For simplicity, we consider plane waves in the simulations, as sketched in Fig. 1(a). The test object consists of three sets of bars of widths and gaps between them of 10 μm, 20 μm and 40 μm, Fig. 1(b). The parameters selected for the simulations are typical for light optical holography: wavelength $\lambda = 500$ nm, sample-to-detector distance 40 mm, sampling is $512 \times 512$ pixels, pixel size $10 \times 10$ μm$^2$ as shown in Fig. 1(c). The reconstruction of the hologram, depicted in Fig. 1(d), shows that all the bars are resolved. The simulation and reconstruction of the hologram are obtained by using the angular spectrum algorithm [17].

Next, we introduce vibrations and assume that the vibration function is described by a Gaussian distribution:

$$V(x, y) = G(x, y) = \exp\left(-\frac{x^2 + y^2}{2\sigma^2}\right). \tag{8}$$

The blurred hologram and its reconstruction for $\sigma = 10$ μm and for $\sigma = 20$ μm are shown in Fig. 1(e) – (h) and Fig. 1(i) – (l), respectively. The degradation of both contrast and extent of the interference pattern associated with the blurred hologram is apparent in Fig. 1(e) and (i). As a result, the reconstructions are also blurred and not all sample bars are resolved, Fig. 1(f) and (j). Deblurred holograms, shown in Fig. 1(g) and (k), were obtained by applying Eq. (5) where $\text{FT}[V(X,Y)]$ was analytically calculated and $\beta = 0.01$. The resulting reconstruction demonstrate that the sample bars appear to be better resolved, Fig. 1(h) and (l).

When $\sigma = 10$ μm, which corresponds to the width and the gap between the thinnest bars in the sample, those bars are resolved in the reconstruction only after the deblurring procedure has been applied, Fig. 1(g) – (h). When $\sigma = 20$ μm, which corresponds to the width and the separation of the second smallest bars in the sample (20 μm), again those bars (the second smallest bars) are resolved in the reconstruction only after having applied the deblurring procedure. However the smallest bars of 10 μm remain unresolved, Fig. 1(k) – (l). This observation leads to the following conclusions. (1) Even if the vibration function is known, deconvolution cannot completely compensate for the vibrations but it can enhance the resolution. (2) The final resolution is proportional to the extent of the vibrations. For example, vibrations which are Gaussian distributed with a standard deviation $\sigma$ can be suppressed by a deblurring deconvolution and provide reconstructions where features as small as $\sigma$ can be resolved.

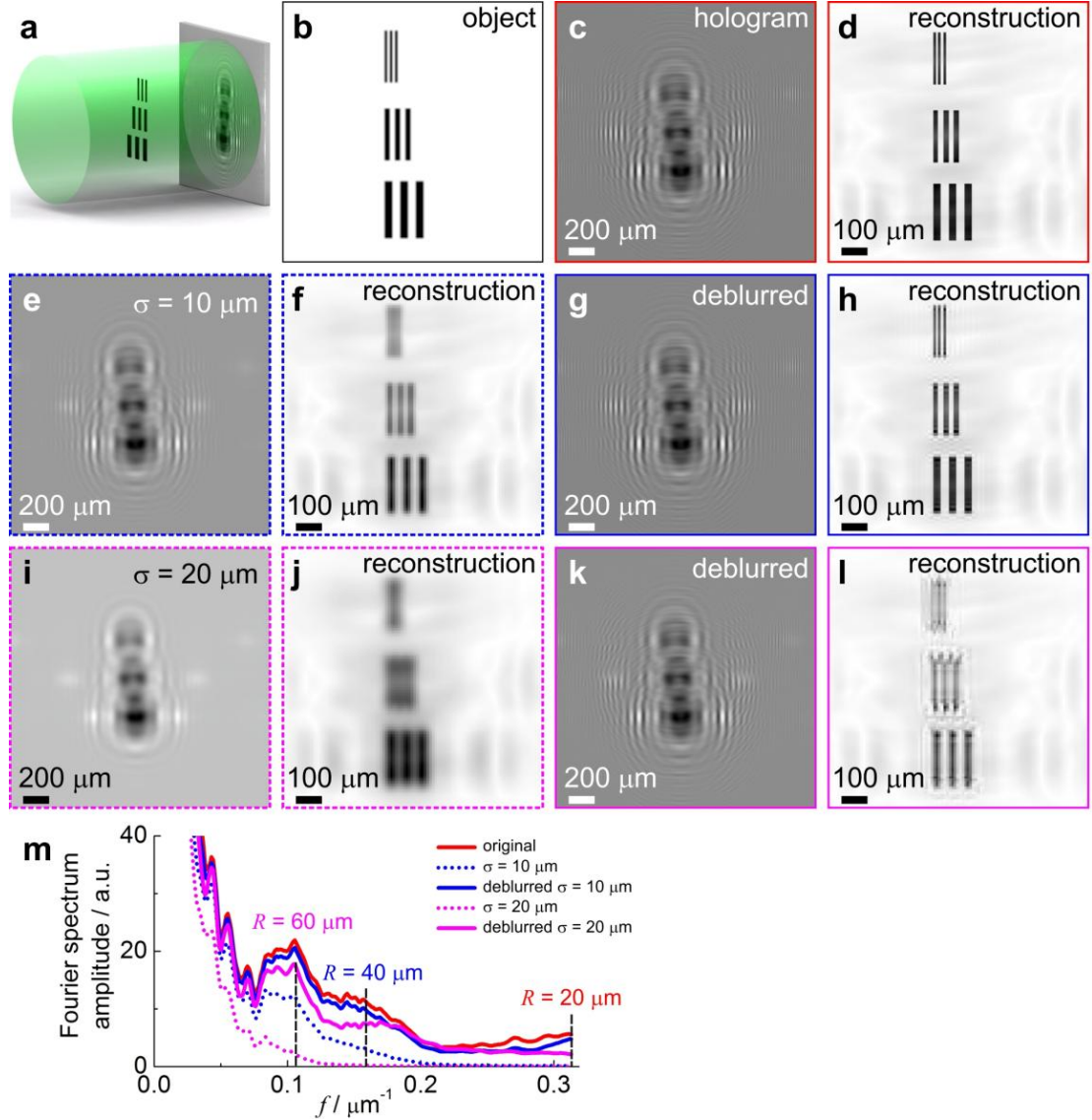

Fig. 1. Simulated example of vibration compensation in an in-line hologram. (a) Sketch of the experimental arrangement. (b) Test object consisting of three sets of bars exhibiting widths and gaps between them of 10 µm, 20 µm and 40 µm. (c) In-line hologram simulated for a plane wave of wavelength $\lambda = 500$ nm, sample-to-detector distance 40 mm, sampled with 512 × 512 pixels, pixel size 10 × 10 µm$^2$. (d) Amplitude of the object reconstructed from the hologram shown in (c). (e) Hologram obtained by superposition of 10000 shifted holograms, the shifts are Gaussian distributed with a standard deviation of $\sigma = 10$ µm (corresponding to 1 pixel). (f) Amplitude of the object reconstructed from the hologram shown in (d). (g) Deblurred hologram obtained by deconvolution of the hologram shown in (e) with the $G(x, y)$ function with $\sigma = 10$ µm. (h) Amplitude of the object reconstructed from the hologram shown in (g). (i) Hologram obtained by superposition of 10000 shifted holograms, the shifts are Gaussian distributed with a standard deviation of $\sigma = 20$ µm (2 pixel). (j) Amplitude of the object reconstructed from the hologram in (i). (k) Deblurred hologram obtained by deconvolution of the hologram (i) with the $G(x, y)$ function with $\sigma = 20$ µm. (l) Amplitude of the object reconstructed from the hologram in (k). (m) Radial averaged amplitude of the Fourier spectra (calculated as $\langle |\text{FT}(o)| \rangle_f$ where $\langle ... \rangle_f$ denotes averaging over frequencies $f$) of the object reconstructed from the original, the blurred and deblurred holograms. The cut-off frequencies are calculated with Eq. (11) for a resolution of 20 µm, 40 µm and 60 µm. They amount to: 0.314 µm$^{-1}$, 0.157 µm$^{-1}$ and 0.105 µm$^{-1}$ correspondingly and are marked in the spectra.

### 3.1 Resolution estimation

The ultimate resolution is given by the diffraction limit and solely associated with geometrical parameters:

$$R_{NA} = \frac{\lambda}{2NA} = \frac{\lambda}{2\sin\vartheta} \approx \frac{\lambda}{2\tan\vartheta} = \frac{\lambda z}{S}, \quad (9)$$

where the angle $\vartheta$ is one-half of the angular aperture, $\tan\vartheta = \frac{S}{2z}$, $S$ is the screen size and $z$ is the distance between the sample and the detector. Equation (9) can be re-written as [1] $R_{NA} \approx \frac{\lambda z}{N\Delta_S}$, where $\Delta_S$ is the pixel size of the detector, in which case the diffraction-limited resolution equals to the pixel size in the object plane $R_{NA} = \Delta_0$. From Eq. (9) it is apparent that the diffraction-limited resolution depends only on the hologram size $S$.

In digital holography, however, the resolution depends on the pixel size with which the hologram is sampled. According to the Shannon sampling theorem [19-20], it requires at least 2 samplings per period to correctly represent a periodic signal. This implies that the finest resolved fringe must be sampled with at least 2 pixels, independent of the size of the pixel. This requirement transfers into the following resolution (details of this derivation are provided in [2]):

$$R_S = 2\Delta_0. \quad (10)$$

Thus, the intrinsic resolution of a digital holographic system is given by the largest of the two resolutions $R_{NA}$ or $R_S$.

In practice, the resolution can be evaluated from the spectrum of the hologram (or reconstructed object) [2] as the highest frequency where the spectrum is non-zero $f_{max}$, where $f = \sqrt{\mu^2 + \nu^2}$. The resolution corresponding to the highest frequency is then given by:

$$R_f = \frac{2\pi}{f_{max}}. \quad (11)$$

In the spectrum of the object (or hologram), the frequency which corresponds to the finest fringes with a period of 2 pixel is found at (N/period of structure)=N pixel/2 pixel=N/2. This is the highest possible frequency in the spectrum and it equals to

$$\left(f_{max}\right)_{max} = \Delta_f \frac{N}{2} = \frac{2\pi}{N\Delta_0} \frac{N}{2} = \frac{\pi}{\Delta_0}, \quad (12)$$

where $\Delta_f$ is the pixel size in the Fourier domain. The corresponding best possible resolution, calculated using Eq. (11) is the same as the resolution defined by the sampling requirement:

$$\left(R_f\right)_{max} = \frac{2\pi}{\left(f_{max}\right)_{max}} = 2\Delta_0. \quad (13)$$

For example, for our sample consisting of bars, the smallest bars are only 1 pixel (10 μm) in width and the gap between the bars amounts to 1 pixel (10 μm). These bars can be reconstructed from the original hologram. However, it is important to note that the resolution here is not 1 pixel (10 μm), but

$R_S = (R_f)_{max} = 2\Delta_0 = 2$ pixels (20 μm). This is obvious from the fact that although the gap between the bars is 10 μm, the separation between the bars which is counted as the center-to-center distance, is 20 μm. Thus, the resolution of the original hologram amounts to 20 μm.

From visual inspection of the holograms and their reconstructions shown in Fig. 1, the following conclusions can be drawn. The resolution of the hologram blurred by vibrations with $\sigma = 10$ μm is 40 μm because only the middle set of bars can be resolved. The resolution of the hologram blurred with $\sigma = 20$ μm is 60 μm because only the bottom set of bars can be resolved. The cut-off frequencies are found in the curves shown in Fig. 1(m) at approximately 0.24 μm$^{-1}$ and 0.13 μm$^{-1}$, for blurring $\sigma = 10$ μm and $\sigma = 20$ μm, respectively. These values according to Eq. (11) provide a resolution of 26 μm and 48 μm, which is slightly better than the resolution derived by visually inspection of the reconstructed objects.

## 3.2 Selecting parameters for optimal deblurring

When the parameters of the vibration function (as for example $\sigma$ for a Gaussian function) are not known, one can perform deblurring of holograms at a sequence of different parameters and then select optimal parameters at which the most resolution enhanced reconstruction is achieved, as illustrated in Fig. 2. Figure 2 demonstrates that when a hologram is blurred with a Gaussian vibration function with $\sigma = 10$ μm, and then it is subsequently deblurred with a Gaussian vibration function with different $\sigma$, the best reconstruction is achieved when $\sigma = 10$ μm.

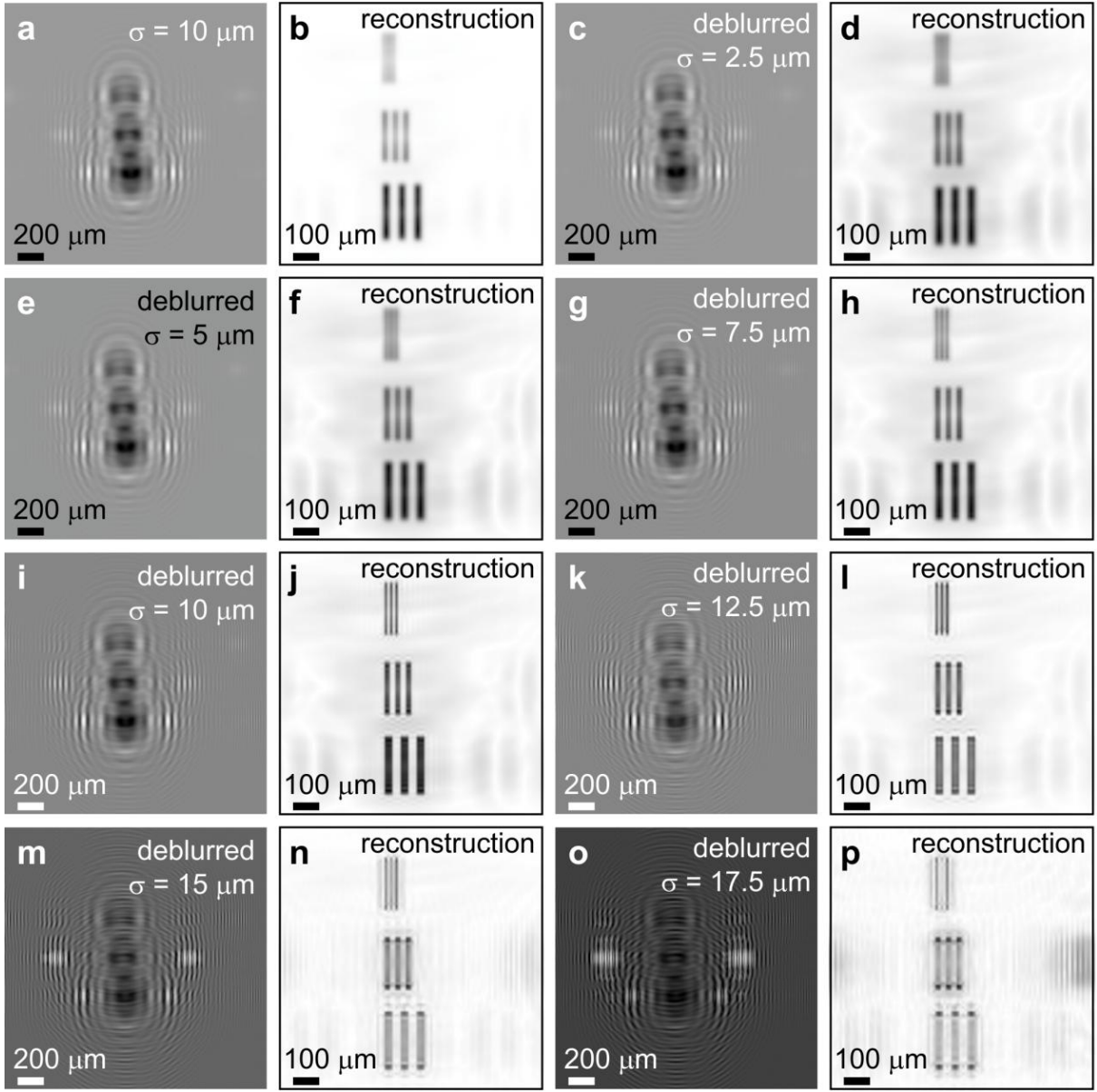

Fig. 2. Deblurring with tuning the deblurring parameter $\sigma$. (a) In-line hologram simulated for a plane wave of wavelength $\lambda = 500$ nm, sample-to-detector distance 40 mm, sampled with 512 × 512 pixels, pixel size 10 × 10 µm². The blurred hologram is obtained by a superposition of 10000 shifted holograms, the shifts are Gaussian distributed with a standard deviation of $\sigma = 10$ µm (corresponding to 1 pixel). (b) Amplitude of the object reconstructed from the hologram shown in (d). The next images are shown in pairs, left: deblurred hologram and right: amplitude of the reconstructed object. (c) – (d) $\sigma = 2.5$ µm, (e) – (f) $\sigma = 5$ µm, (g) – (h) $\sigma = 7.5$ µm, (i) – (j) $\sigma = 10$ µm, (k) – (l) $\sigma = 12.5$ µm, (m) – (n) $\sigma = 15$ µm, (o) – (p) $\sigma = 17.5$ µm.

### 3.3 Note on axial vibrations

Until now we addressed only lateral vibrations limited to the $(x, y)$-plane. Of course, vibrations along the $z$-axis can occur as well. Changes in $z$-distance mainly change the magnification of the interference pattern, which can be illustrated as following. For example, for a hologram recorded with a plane wave, the wave scattered by the object at an angle $\vartheta$ arrives at the detector at the $z \tan \vartheta$ position (where $z$ is the sample-to-detector distance, or the reconstruction distance). This position linearly scales with the z-distance. For a hologram recorded with a spherical wave, the magnification factor is given by $D/z$ (here $D$ is the source-to-detector distance and $z$ is the source-to-sample distance) and thus the magnification here also scales with the $z$-distance.

A hologram obtained during axial vibrations can be represented as a sum of holograms at different magnifications

$$H(X,Y) = \sum_i H_0(M_i X, M_i Y). \tag{14}$$

The resulting hologram is a blurred hologram, see simulations shown in Fig. 3. Unfortunately, Eq. (14) cannot be represented in form of a convolution since a convolution does not change the size of an image. Thus, even if the scaling factors $M_i$ were known, an accurate deblurring procedure would not be possible. However, to some approximation rescaling of an image can be approximated with shifts of its parts. For example, for a single pixel located at $(x_0, y_0)$, its position after rescaling will be $(Mx_0, My_0)$. The same position would be achieved as a result of a lateral shift by $((M-1)x_0, (M-1)y_0)$. Because such shift depends on the coordinates of the pixel a precise deconvolution kernel cannot be derived. However, trying out a deconvolution with a Gaussian distributed kernel with different $\sigma$ can result in a somewhat deblurred reconstruction as illustrated in Fig. 3.

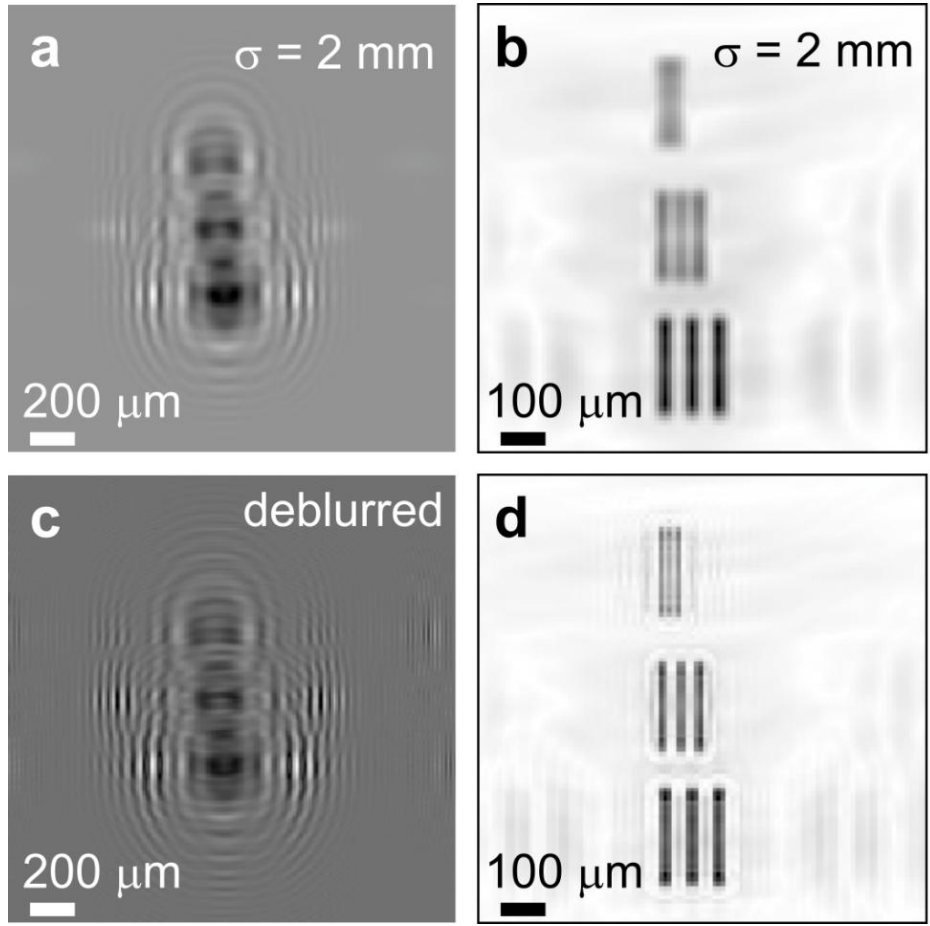

Fig. 3. Simulated example of axial vibration compensation in an in-line hologram. (a) In-line hologram simulated for a plane wave of wavelength $\lambda = 500$ nm, sample-to-detector distance 40 mm, sampled with $512 \times 512$ pixels, pixel size $10 \times 10$ µm$^2$. The hologram is obtained by superposition of 10000 holograms of a $z$-shifted object, the shifts are Gaussian distributed with a standard deviation of $\sigma = 2$ mm (2 pixel). (b) Amplitude of the object reconstructed from the hologram in (b). (c) Hologram obtained by deconvolution of the hologram in (a) with a $G(x, y)$ function with $\sigma = 14$ µm. (d) Amplitude of the object reconstructed from the hologram in (c).

## 3.4 Directional motion blur

The method of deblurring can be also illustrated for another type of motion that is different from vibrations distributed in form of a Gaussian function. For example, the hologram could be moving in one direction during the acquisition. A hologram which is blurred by this type of directional motion is simulated and shown in Fig. 4(a). The motion path here is a horizontal line whose length is almost half of the central part of the hologram, shown in Fig. 4(b). The object reconstructed from the blurred hologram does not exhibit any sharp features, as evident from Fig. 4(c). The deblurred hologram is calculated according to Eq. (5) where $V(X,Y)$ is the motion path distribution (shown in Fig. 4(b)) and $\beta = 1$. The deblurred hologram is shown in Fig. 4(d). The corresponding reconstructed object (shown in Fig. 4(e)) exhibits all the bars resolved. This example illustrates that provided the blurring function is known, the resolution can be restored almost to the original resolution in the absence of any blurring.

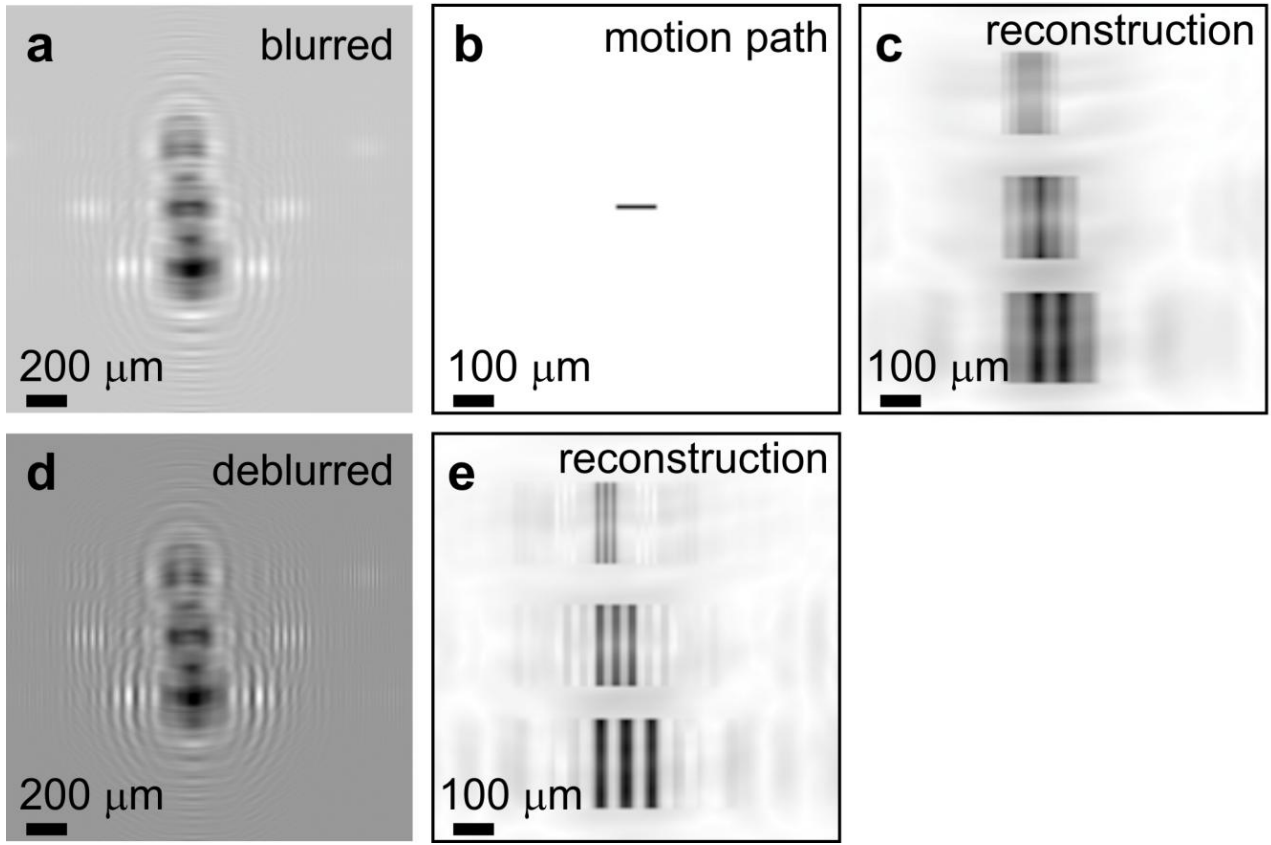

Fig. 4. Simulated example of a hologram blurred by a directional motion. (a) In-line hologram simulated for a plane wave of wavelength $\lambda = 500$ nm, sample-to-detector distance 40 mm, sampled with 512 × 512 pixels, pixel size 10 × 10 µm². The blurred hologram is obtained by the superposition of 10 holograms that were shifted horizontally at a step of 10 µm, the motion path is shown in (b). (c) Amplitude of the object reconstructed from the blurred hologram shown in (a). (d) Hologram obtained by deblurring of the hologram shown in (a). (e) Amplitude of the object reconstructed from the deblurred hologram shown in (d).

## 4. Experimental results

### 4.1 Optical holograms

Conventionally, light optical holograms can be acquired without significant vibrations. In practice, the resolution expected from a light optical hologram can be as small as twice the wavelength. For example, microspheres of 1 um in diameter can be imaged by light optical holography employing a wavelength of 532 nm [21]. Vibrations of the order of one micron or less thus should not severely reduce the contrast of the interference fringes in a light optical hologram. However, still the resolution can be somewhat enhanced by deblurring an optical hologram as we now show below.

An example of a light optical hologram is shown in Fig. 5. The experimental setup is sketched in Fig. 5(a), the hologram was acquired with a divergent spherical wave and a scanning electron micrograph of the test object is shown in Fig. 5(b). We assumed that vibration amplitudes were of the of order of 1 μm and performed a deconvolution of the original hologram as described by Eq. (5) ($\beta = 0.5$) with a Gaussian with $\sigma = 1$ μm as described by Eq. (8). From comparing the hologram before and after deconvolution, Fig. 5(c) and (d), we can see that the contrast of the interference fringes is enhanced. The spectrum exhibits more high-frequency components after deblurring, see Fig. 5(g). The diffraction-limited resolution of the hologram estimated with Eq. (9) is $R_{NA} = 0.89$ μm, and the sampling resolution estimated with Eq. (10) is $R_S = 2\Delta_0 = 308.4$ nm. Thus, the intrinsic resolution amounts to $R_{NA} = 0.89$ μm. The resolution of the original hologram estimated from the spectrum of the reconstructed object (Eq. (11)) amounts to $R_f \approx \frac{2\pi}{3.3 \ \mu m^{-1}} = 1.9$ μm and the resolution of the deblurred hologram amounts to $R_f \approx \frac{2\pi}{4 \ \mu m^{-1}} = 1.6$ μm. Although this enhancement is insignificant, the visual appearance of the reconstruction appears much sharper, as can be seen from comparing Fig. 5(e) and Fig. 5(f). Moreover, the profiles along the finest features of the object display sharper contours in the reconstruction obtained from the deblurred hologram, Fig. 5(h).

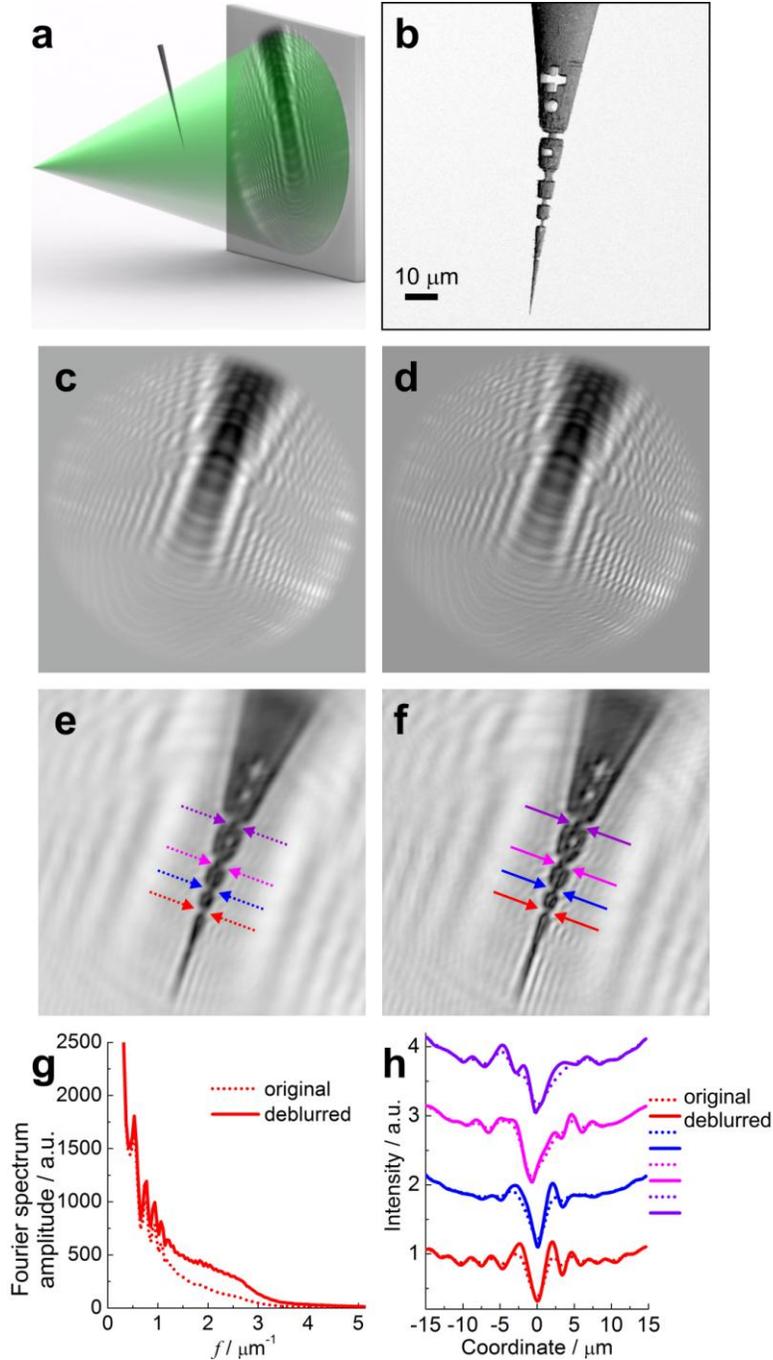

Fig. 5. Light optical hologram of a tungsten tip with a structure milled with a focused ion beam. (a) Sketch of the experimental arrangement. (b) Scanning electron micrograph of the micro-structured tip. (c) Light optical in-line hologram recorded with a divergent wave. The parameters for the hologram acquisition are: the source-to-detector distance is 5cm, the hologram size is 30 × 30 mm$^2$ and sampled with 1000 × 1000 pixels, wavelength = 532 nm, the source-to-sample distance is obtained during reconstruction as the distance where the object appears in focus and it amounts to 257 μm. The hologram was normalized by division with the background image, that is obtained under the very same conditions but without the object being present [21-22]. (d) Hologram after deblurring with $G(x, y)$ with $\sigma = 1$ μm as described by Eq. (8). (e) Object reconstructed from the original hologram and (f) from deblurred hologram. (g) Radial profile of the amplitude of the spectrum of the objects reconstructed from the two holograms. (h) Profiles along the colored arrows in the two reconstructions.

## 4.2 Electron holograms

The problem of vibrations is severe in in-line holography with electrons [23-27] whereby the employed wavelengths are very short, in the Angstrom regime. The resolution in such holograms can intrinsically be as small as one Angstrom or less, and atomic resolution should in principle be possible. However, one of the reasons that cause a degraded resolution are residual mechanical vibrations of either the source or the object. Even vibrations of the order of the inter-atomic distance can smear out the fine fringes in the interference pattern that carry the atomic resolution information.

An example of resolution enhancement in a low-energy electron hologram of a bundle of single-walled nano-tubes (SWNT) stretched over a hole in a carbon film is shown in Fig. 6. A schematic of the experimental setup is shown in Fig. 6(a) and details of the experimental setup are provided elsewhere [23]; here the source of the coherent electron wave is a sharp tungsten tip [28-29]. 20 holograms were recorded with 220 eV kinetic energy electrons (wavelength = 0.83 Å), aligned and averaged to compensate for systematic drift [8], the resulting hologram is shown in Fig. 6(b). The hologram was deblurred by deconvolution with $G(x, y)$ with $\sigma = 1.5$ nm as described by Eq. (8). The same hologram after deblurring exhibit much more finer interference fringes (Fig. 6(c)), which can also be seen by comparing the magnified regions of both holograms shown in Fig. 6(d) and Fig. 6(e). The Fourier spectrum of the deblurred hologram exhibit more high-frequency components which provide high-resolution information, shown in Fig. 6(f). As a result, the reconstruction of the deblurred hologram appears sharper and the contour of the SWNT bundle shows finer details, Fig. 6(g). The resolution of the original hologram estimated from the spectrum of the reconstructed object (Eq. (11)) amounts to $R_f \approx \dfrac{2\pi}{2 \text{ nm}^{-1}} = 3.1$ nm and the resolution of the deblurred hologram amounts to $R_f \approx \dfrac{2\pi}{3 \text{ nm}^{-1}} = 2.1$ nm. The reconstructions here were obtained by single-sideband holography [30] which allows obtaining twin-image free reconstructions on one side of the object [31].

Another example of a low-energy electron hologram is that of a single Bovine Serum Albumin (BSA) protein deposited on free-standing graphene and shown in Fig. 7. Previously reported hologram and reconstruction of the BSA protein [32] can be enhanced by our method. As a result, features which were blurred in the previous reconstruction, can now be distinguished.

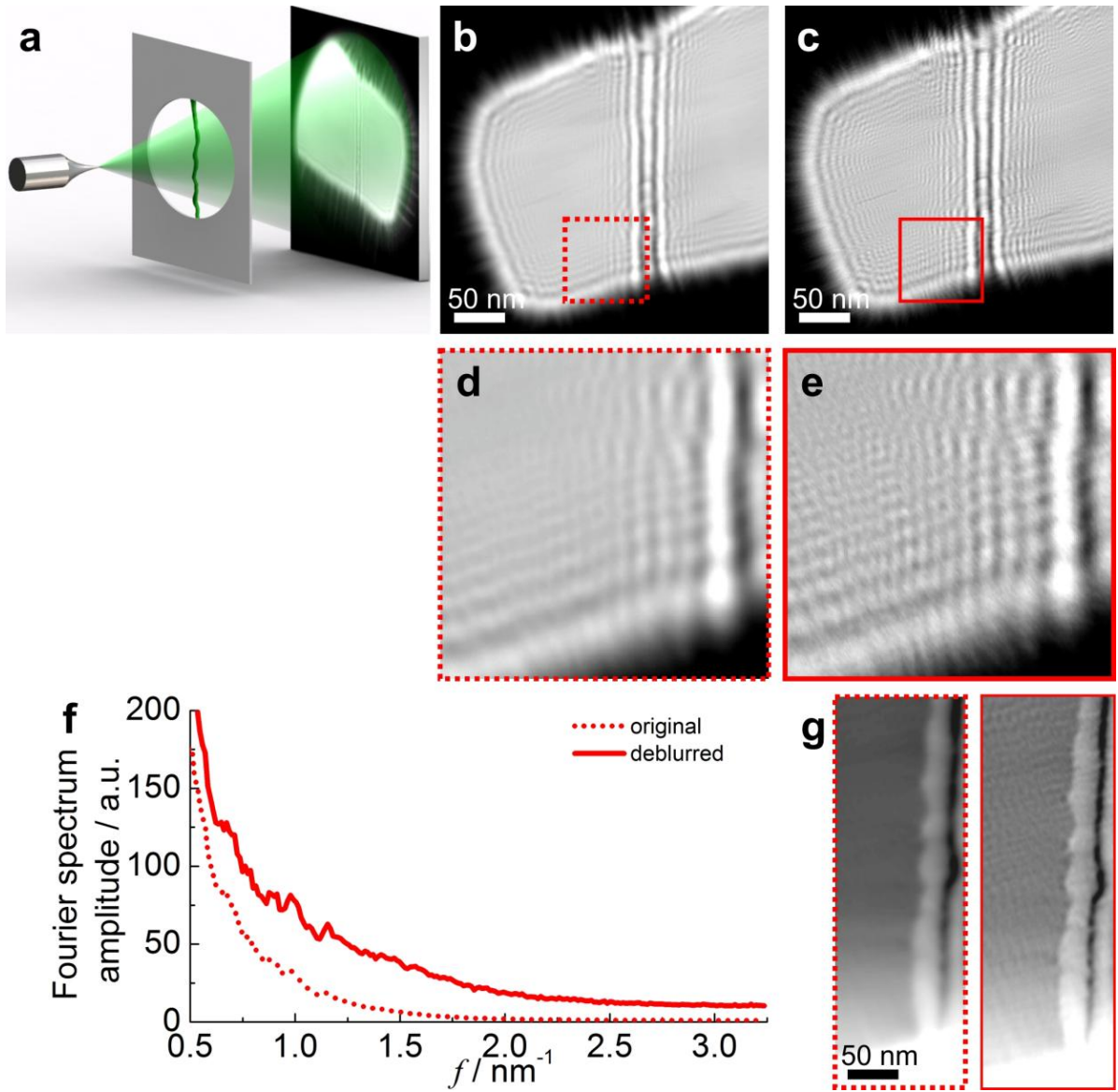

Fig. 6. Low-energy electron hologram of a bundle of single-walled nano-tubes (SWNT). (a) Sketch of the experimental setup. (b) Hologram recorded with electrons of 220 eV kinetic energy (wavelength = 0.08 nm). The other parameters of the acquisition are: the source-to-detector distance is 180 mm, the hologram is 26 × 26 mm² and it is sampled with 1000 × 1000 pixels, the source-to-sample distance is obtained during the reconstruction as the distance where the object appears in focus and amounts to 3.4 μm. The hologram was normalized by division of the original hologram with background image, which was obtained by surface fitting of the patches of the intensity distribution that are free from interference pattern. (c) Hologram (a) after deblurring with $G(x,y)$ with $\sigma = 1.5$ nm as described by Eq. (8). (d) and (e) a magnified region in holograms (b) and (c), respectively. (f) Radial profiles of the amplitude of the spectrum of the objects reconstructed from the two holograms. (g) SWNT reconstructed (left) from the original hologram and (right) from deblurred hologram.

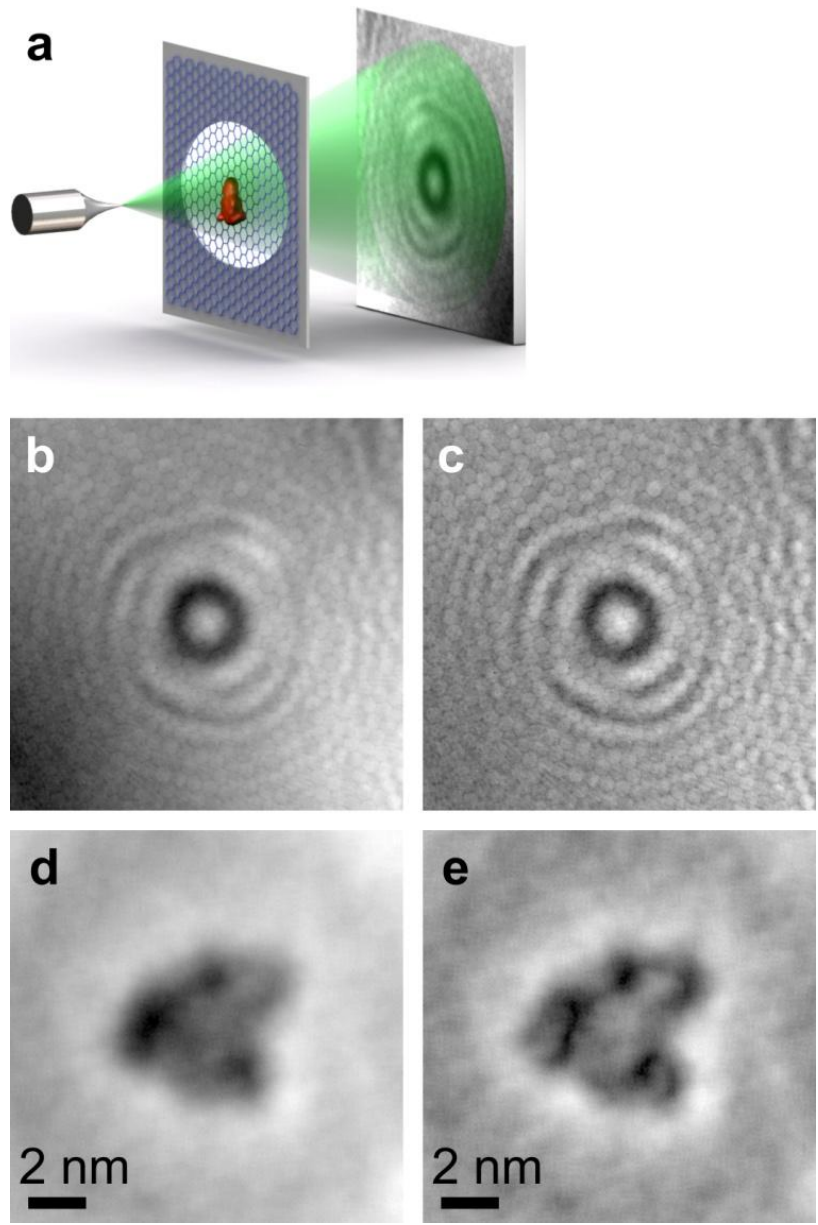

Fig. 7. Enhancement of a low-energy electron hologram of a BSA protein which was previously published [32]. (a) Sketch of the experimental setup. (b) Hologram recorded with electrons of 62 eV kinetic energy (wavelength = 0.16 nm) at the source-to-detector distance of 70 mm, the source-to-sample distance is obtained during the reconstruction as the distance where the object appears in focus and amounts to 250 nm. (c) Hologram after deblurring with $G(x, y)$ with $\sigma = 1.8$ nm as described by Eq. (8). (d) and (e) BSA reconstructed (left) from the original hologram and (right) from deblurred hologram.

## 5. Discussion

We demonstrated resolution enhancement in holography in the case when holograms are blurred due to vibrations during data acquisition. A vibration function is described by a Gaussian distribution which can be assumed in a general case. When performing the deblurring procedure the parameters of the vibration function (as for example $\sigma$ for a Gaussian function) can be varied until a sharp reconstruction is achieved, similar to tuning the focus knob in an optical microscope. When no a priori information is provided about the vibration function, a good starting point for modeling the vibration function is checking the extent of the Fourier spectrum of the hologram and model the initial deconvolution function such that the frequencies that are degraded would be enhanced. Precise knowledge of the vibration function distribution $V(X,Y)$ allows better deblurring of holograms by applying Eq. (5), as we illustrated in the case of a hologram blurred by directional motion. In reality, the vibration function can have any distribution and methods of blind deconvolution [12] (and references therein) can be applied.

The method does not introduce artificial fringes. This can be illustrated with the following example. The Fourier transform of a constant background is described by a $\delta$-function at frequency zero. The deconvolution function has the value 1 at the frequency zero. Thus the product of the Fourier transform of the background with the deconvolution function is again a $\delta$-function at the frequency zero. The inverse Fourier transform again provides the same constant background, no additional fringes appear after deconvolution.

The deblurring procedure enhances the higher frequencies in the Fourier spectrum, which can also enhance the noise. Therefore, the method works best for holograms that are obtained with a long acquisition time. This allows the realization of all possible vibrations and thus a better representation of the vibration function. At the same time, a long acquisition time ensures that the noise is reduced in the acquired hologram.

The method can suppress, but not eliminate completely, blurring caused by object movements in the $(x, y)$-plane and along the $z$-axis. Other sources of blurring not addressed in this paper include, for example, beam intensity noise or partial coherence.

We demonstrated that simple deblurring methods applied to holograms can enhance the contrast and extent of the interference pattern which in turns improves the resolution of the reconstructed objects. The method does not depend on the properties of the object that was used to create the hologram. Thus, any kind of object, be it amplitude or phase objects can subject to this deblurring method.

Although the main message of this work is the deblurring of holograms, we would like to briefly discuss whether deblurring methods can be applied not only to holograms but to the reconstructed objects instead. It can be shown, that the distribution of a hologram can be written as convolution of the object distribution with the propagation factor $H_0(X,Y) \approx o(X,Y) \otimes s(X,Y)$ where $s(X,Y)$ is the propagation Fresnel function [2]. A blurred hologram is then given by $H_0(X,Y) \approx [o(X,Y) \otimes s(X,Y)] \otimes V(X,Y)$ which can be re-written as $H_0(X,Y) \approx [o(X,Y) \otimes V(X,Y)] \otimes s(X,Y)$ which is a hologram of the blurred object. Thus, using this approximation, the hologram can be reconstructed conventionally first and then the deblurring algorithms can be applied to the reconstruction. The object distribution can be complex-valued; that is the object can possess absorption as well as phase properties. The blurred object can be represented as

$$o(x, y) = o_0(x + \Delta x_1, y + \Delta y_1) + o_0(x + \Delta x_2, y + \Delta y_2) + ... =$$
$$o_0(x, y) \otimes \sum_i \delta(x + \Delta x_i, y + \Delta y_i) = o_0(x, y) \otimes V(x, y).$$

From this formula, it is evident that in order to correctly extract the amplitude and the phase of the object, one has to first perform the deblurring of the reconstructed complex-valued $o(x, y)$, and subsequently separate the complex-valued distribution into its amplitude and phase distributions. Applying the deblurring methods to the reconstructed objects would allow using those deblurring algorithms which are based on detection and enhancement of sharp edges of the object [16].

To conclude, we believe that our findings open a new venue for adapting various motion deblurring algorithms developed in photography [15-16] (and references therein) for resolution enhancement in holography.

**References and links**